\documentclass[twocolumn,showpacs,aps,prl,amsmath,amssymb]{revtex4}
\usepackage{graphicx}
\usepackage{dcolumn}
\usepackage{bm}
\usepackage{amsmath,epsfig}
\newcommand{\ket}[1]{\left\vert#1\right\rangle}
\newcommand{\bra}[1]{\left\langle#1\right\vert}
\newcommand{\s}{\uparrow}
\newcommand{\g}{\downarrow}
\begin{document}
\author{F. Ciccarello\mbox{$^{1,2,3}$}}
\author{M. Paternostro\mbox{$^{3}$}}
\author{M. S. Kim\mbox{$^{3}$}}
\author{G. M. Palma\mbox{$^{2}$}}
\affiliation{\mbox{$^{1}$} CNISM and Dipartimento di Fisica e
Tecnologie Relative, Universit\`{a} degli Studi di Palermo, Viale
delle
Scienze, Edificio 18, I-90128 Palermo, Italy \\
\mbox{${\ }^{2}$} NEST-INFM (CNR) \& Dipartimento di Scienze
Fisiche ed Astronomiche, Universit\`{a}
degli Studi di Palermo, Via Archirafi 36, I-90123 Palermo, Italy\\
\mbox{${\ }^{3}$} School of Mathematics and Physics, Queen's
University, Belfast BT7 1NN, United Kingdom}

\begin{abstract}
We present a scheme for the extraction of singlet states of two
remote particles of arbitrary quantum spin number. The goal is
achieved through post-selection of the state of interaction
mediators sent in succession. A small number of iterations is
sufficient to make the scheme effective. We propose two suitable
experimental setups where the protocol can be implemented.
\end{abstract}


\title{Extraction of singlet states from non-interacting high-dimensional
spins}

\maketitle

Achieving control at the quantum level is a pivotal requirement for
the grounding of quantum technology and the development of reliable
protocols for information processing. Frequently, state-manipulation
of a quantum device needs the connection of remote nodes of a
network and the creation of their entangled state. Such a {\it
delocalized architecture} has received strong experimental
attention, especially at the quantum optics level. Heralded
entanglement of remote atomic ensembles or individually-trapped ions
has been produced and atom-photon entanglement has been
observed~\cite{varie1}. The transfer of prebuilt entanglement to
distant systems has been proposed as a way to distribute quantum
channels~\cite{varie2}.

A different approach exploits a mediated interaction between two
remote nodes, $1$ and $2$, by means of their sequential coupling to
the same ancillary system $e$: The ancilla can bring to system $2$
the information that has been previously impressed on it by its
interaction with system $1$. Recently, this idea has been used in a
solid-state context involving multiple electron scattering between
magnetic impurities~ \cite{yang-yasser-giorgi,ciccarello,nakazato}.
Interestingly, $e$ can also be used so as to condition the state of
$1$ and $2$. {Once a three-body correlated state is established by
means of bilocal $1-e$ and $2-e$ interactions}, by measuring the
state of $e$ we could project the remote systems onto entangled
states with a non-zero
probability~\cite{yang-yasser-giorgi,ciccarello,nakazato,smer}. In
these examples, $1$ and $2$ are embodied by two-level systems whose
finite Hilbert space bounds the entanglement that can be
shared~\cite{mauro-prl}. Overcoming such a limitation is an
important task deserving attention.

Here we present a scheme that allows the ``extraction'' of maximally
entangled states via an effective {non-demolition} Bell measurement
performed onto the state of two spin-$s$ particles. This occurs
through repeated injection and post-selection of simple mediators,
each undergoing multiple scattering and spin-flipping between the
two spins~\cite{daniel}. Besides achieving the maximum number of
ebits allowed to two spin-$s$ systems, the protocol provides a
procedure for accumulating entanglement. Remarkably, our protocol
does not require interaction-time tuning. In our scheme maximal
entanglement is stable against the parameters of the conditioned
dynamics, which is a clear advantage in experimental
implementations. In order to fix the ideas, we first describe the
protocol {in terms of} a system composed of a conduction electron
and two magnetic impurities. This will allow us to clearly
{illustrate} the relevant features of our scheme. Later, we show how
a cavity-quantum electrodynamics (QED) system, consisting of two
multilevel atoms interacting with a photon field, can also embody
the desired dynamics and allows a prompt experimental implementation.
\begin{figure}[b]
\centerline{\includegraphics[width=0.28\textwidth]{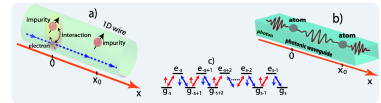}}
\caption{{(Color online) Setups for the implementation of
our scheme in nanowire {\bf (a)} and cavity-QED {\bf (b)}. {\bf (c)}
Multilevel atom embodying a spin-$s$ particle for the setup in panel
{(b)}, where symbols $\s,\g$ indicate, abstractly, proper
polarization of a photon.}} \label{Fig1}
\end{figure}
We consider a quasi one-dimensional (1D) wire, such as a
semiconductor quantum wire~\cite{datta} or a single-wall carbon
nanotube~\cite{nanotube}, where two identical spin-$s$ magnetic
impurities 1 and 2 are embedded at positions $x_{1}\!=\!0$ and
$x_{2}\!=\!x_0$ [see Fig.~1(a)]. Left-incident single electrons undergo
multiple scattering between the two impurities {and simultaneous
spin-flipping}. Assuming that the electron's coherence length
exceeds $x_0$ and that each electron occupies only the lowest
subband, the Hamiltonian reads (we set $\hbar\!=\!1$)
$\hat{H}\!=\!{\hat{p}^{2}}/({2m^*})\!+\!J \,
\hat{\mbox{\boldmath$\sigma$}}\cdot
[\hat{\mathbf{S}}_{1}\,\delta(x)\!+\!\hat{\mathbf{S}}_{2}\,\delta(x\!-\!x_{0})]$.
Here, $\hat{p}\!=\!-i \nabla,\,m^*$ and $\hat{\bm{\sigma}}$ are the
electron momentum, effective mass and Pauli spin operator
respectively. $\hat{\mathbf{S}}_{i}$ is the spin-$s$ operator of the
 impurity $i\!=\!1,2$ and $J$ is the Heisenberg exchange
coupling constant whose dimensions are frequency times length. Due
to the elastic nature of the interactions, the energy spectrum reads
$E\!=\!k^2/2m^*$ ($k$ is the electron wavevector). We label with
$\hat{\mathbf{S}}=\hat{\bm{\sigma}}+\hat{\mathbf{S}}_{1}\!+\!\hat{\mathbf{S}}_{2}$
the total spin of the system, while $m_i$ and $m_e\!=\!\pm1/2$ are the
quantum numbers associated with $\hat{S}_{iz}$ and $\hat{\sigma}_z$,
respectively. From now on, we denote $\{1/2,-1/2\}$ by $\{\s,\,\g\}$
and, for convenience, we use the basis of product states
$\ket{m_e,\{m_i\}}\!=\!\ket{m_e}_e\ket{m_1,m_2}_{12}$. {We prepare the
impurities in $\ket{\{m'_i\}}_{12}$. An incoming electron of
wavevector $k$ and spin state $\ket{m'_{e}}_e$ is reflected
(transmitted) in the state $\ket{m_e}_e$, while the impurities' spin
state changes into $\ket{\{m_i\}}_{12}$ with probability amplitude
$r$ ($t$) (we omit the dependence of $r$ and $t$ on $m_{e (i)}$ and
$m'_{e (i)}$). As $\hat{S}_z$ is a constant of motion,
the only non-zero amplitudes are those obeying the selection rule
$m'_{12}+m'_e=m_{12}+m_e$ with $m_{12}\!=\!m_1\!+\!m_2$.} We solve this
scattering problem by finding the steady states
$\ket{k,m'_e,\{m'_i\}}$ with input part $\langle
x\ket{k,m'_e,\{m'_i\}}_{in}\!=\!e^{ikx}\theta(-x)\ket{m'_e,\{m'_i\}}$,
where $\theta(x)$ is the Heaviside step function. Their output part
reads $\langle
x\ket{k,m'_e,\{m'_i\}}_{out}\!=\!\sum_{\alpha=r,t}\langle
x\ket{k,m'_e,\{m'_i\}}_\alpha$ with $\langle
x\ket{k,m'_e,\{m'_i\}}_{\alpha}\!=\!\sum_{m_e,\{m_i\}}\!\alpha
f_{\alpha}(x)\ket{m_e,\{m_i\}}$ and $f_{\alpha}(x)\!=\!e^{i \eta_\alpha{k}x} \theta\left(\eta_\alpha
x\!-\!\frac{1\!+\!\eta_{\alpha}}{2}
 x_0\right)$  ($\eta_r\!=\!-\eta_t\!=\!-1$).
The steady states are computed {at all orders in $J$} solving the
time-independent Schr\"{o}dinger equation and imposing the matching
of the wavefunction at $x_i$'s~\cite{ciccarello}. {We now derive how
an (in general mixed) initial state of the impurities $\rho_{12}$ is
transformed after scattering of an electron incoming in an arbitrary
statistical mixture $\rho_e$ of the spin states $\ket{\s}_e$ and
$\ket{\g}_e$. To this aim,} we consider the state having
$\ket{k}\bra{k}\rho_e\rho_{12}$ as input part, where $\langle
x\ket{k}\!=\!e^{ikx}\theta(-x)$. The output part of such state is found
by expanding it in the basis $\{\ket{k,m'_e,\{m'_i\}}\}$ and
replacing each component of this expansion with the corresponding
output part. A further projection onto the electron's position
eigenstates far from the impurities $\ket{x_r}$ and $\ket{x_t}$
($x_r\!\ll\! 0$, $x_t\!\gg\!x_0$) yields $\sum_{\alpha\!=\!r,t}\langle
x_\alpha\ket{k,m'_e,\{m'_i\}}_{\alpha}\!\bra{k,m'_e,\{m'_i\}}x_\alpha\rangle
\ket{x_{\alpha}}\!\bra{x_{\alpha}}$. After tracing over the
electron's degrees of freedom, the impurities' state becomes
\begin{equation}\label{map_up}
\mathcal{E}_{\rho_e}(\rho_{12})\!=\!\sum_{\mu,\nu\!=\!\s,\g}{\rho_e}_{\nu\nu}(\hat{R}^{\mu}_{\nu}\rho_{12}\hat{R}_{\nu}^{\mu\,\dag}\!+\!\hat{T}^{\mu}_{\nu}\rho_{12}\hat{T}_{\nu}^{\mu\,\dag}),
\end{equation}
where
$\sum_\mu(\hat{R}_{\nu}^{\mu\,\dag}\hat{R}^{\mu}_{\nu}+\hat{T}_{\nu}^{\mu\,\dag}\hat{T}^{\mu}_{\nu})=\openone_{12}$.
Each Kraus operator ${R}_{\nu}^{\mu}$ (${T}_{\nu}^{\mu}$) {depends
only {on $r$'s ($t$'s)} and is physically interpreted as the effect
on $\rho_{12}$ {due to the detection in spin-state $\ket{\mu}_e$ of
a reflected (transmitted) electron incoming in state $\ket{\nu}_e$.
We want to show that, conditioning the map in Eq.~(\ref{map_up}) and
iterating it for $n$ electrons (injected in succession in the same
spin state), singlet-state extraction is efficiently performed. To
achieve this, we first describe what is induced by post-selecting
the state of $n=1$ scattered electrons. Preparation and
post-selection of a given electron spin state, say $\ket{\s}_e$, can
be accomplished using spin-filtering contacts at the input/output
ports of the wire~\cite{spintronics}, each selecting the same spin
state. We obtain the final impurities' state
$\varrho^{(1)}_{12}\!=\!\mathcal{E}_{\s\s}(\rho_{12})\!=\!{(\hat{R}^{\s}_{\s}\rho_{12}\hat{R}_{\s}^{\s\dag}+\hat{T}^{\s}_{\s}\rho_{12}\hat{T}_{\s}^{\s\dag})}/{P_{\s\s}^{(1)}(\rho_{12})}$
with success probability
$P_{\s\s}^{(1)}(\rho_{12})\!=\!\mathrm{Tr}_{12}(\hat{R}^{\s}_{\s}\rho_{12}\hat{R}_{\s}^{\s\dag}\!+\!\hat{T}^{\s}_{\s}\rho_{12}\hat{T}_{\s}^{\s\dag})$.
The state $\varrho^{(n)}_{12}$ corresponding to $n$ electrons being
prepared and post-selected in $\ket{\s}_e$ is obtained
as $\varrho^{(n)}_{12}=\mathcal{E}_{\s\s}^n(\rho_{12})$ with
conditional probability $P_{\s\s}^{(n\ge1)}(\rho_{12})=\prod_{j=1,n}
P_{\s\s}(\varrho^{(j-1)}_{12})$ and
$\varrho^{(0)}_{12}=\rho_{12}$~\cite{marsiglio}.} Here, the rate of
electron-injection is chosen so that, as an electron reaches the
impurities, the previous one has been already scattered off.
\begin{figure}[b]
\centerline{\includegraphics[width=0.17\textwidth]{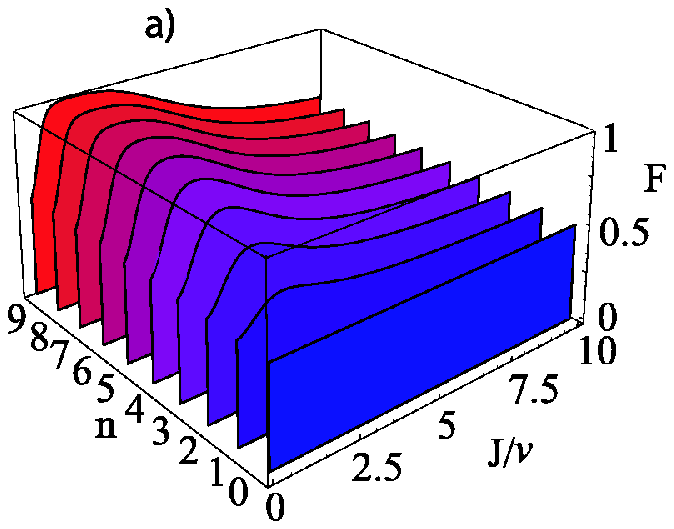}\hskip0.1cm\includegraphics[width=0.17\textwidth]{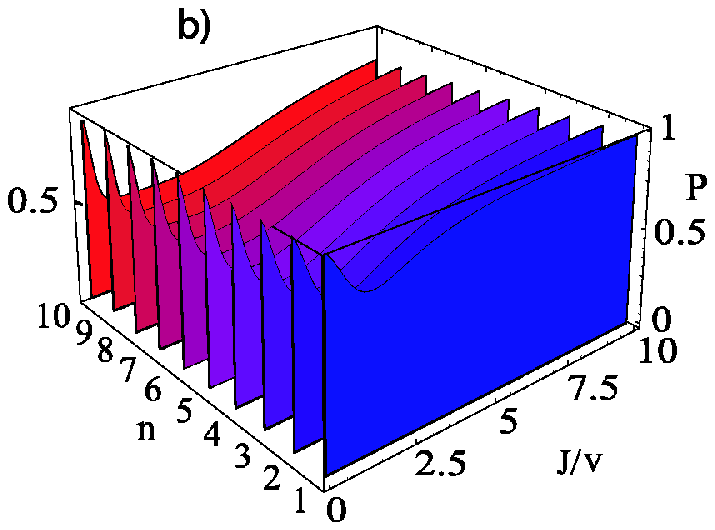}\hskip0.1cm\includegraphics[width=0.17\textwidth]{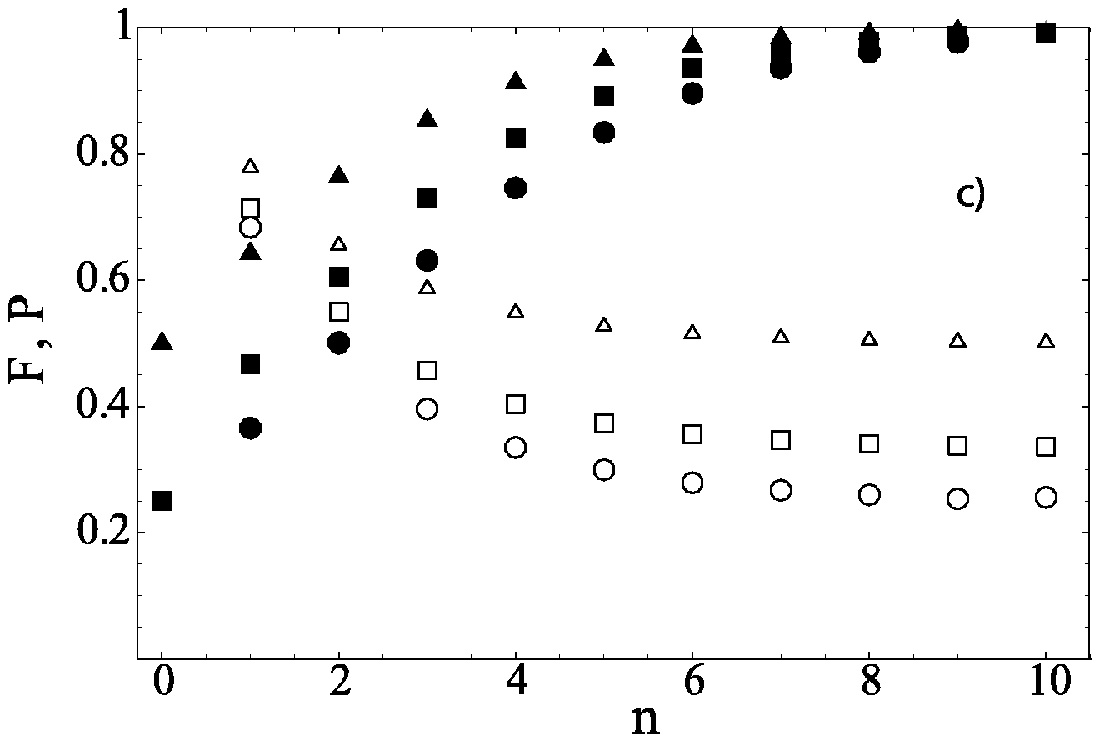}}
\caption{(Color online) (a) and (b) Fidelity and success probability
vs. $J/v$ and $n$ for $s=1/2$. (c) $F^{(n)}$ (filled symbols) and
$P_{\s\s}^{(n)}$ (empty symbols) vs. $n$ for $s=1/2$ and ${J}/v=1.5$
($\blacktriangle,\,\vartriangle$), $s=1$ and ${J/v}=1.2$
($\blacksquare,\,\square$) and $s=3/2$ and ${J/v=1.1}$
($\bullet,\,\circ$) at $kx_0/\pi\in{\mathbb Z}$ ($J/v$ is opimized
for each $s$).} \label{Fig2}
\end{figure}
Let $\ket{\Psi_{s}^-}$ be the singlet state of two spin-$s$
impurities. Using resonance conditions (i.e. $kx_0/\pi\in
\mathbb{Z}$), in Fig.~2(a) and (b) {we consider the case $s\!=\!1/2$ and
plot} the fidelity $F^{(n)}$ of $\varrho^{(n)}_{12}$ with respect to
the singlet $|{\Psi_{1/2}^-}\rangle$ together with
$P_{\s\s}^{(n)}$ as functions of $n$ and $J/v$ for the {initial
product state $\ket{1/2,-1/2}_{12}$} ($v\!=\!k/m^*$ is the electronic
group velocity). Clearly, $F^{(n)}\rightarrow{1}$ for a range of
values {around $J/v\simeq 1.5$} that becomes a plateau when $n$
increases ($n<7$ iterations are enough to get fidelity higher than
0.95). For a fixed value of $J/v$, such convergence is
\emph{exponential} in $n$. Remarkably, although our protocol is
conditioned on the outcomes of $n$ projective measurements all with
the same outcome, the probability of success converges exponentially
to $0.5$. Differently
from~\cite{yang-yasser-giorgi,ciccarello,nakazato}, the scheme is
still {efficient} for a non-optimal $J/v$. Only a larger $n$ is
required, for a fixed $s$. Moreover, the process is robust against
discrepancies of $k$ with respect to resonance conditions and the
use of a stream of mediators with mutually different wavevectors.
{In fact, by considering a Gaussian distribution of wave vectors
centered at $k$ with variance $\sigma$, we have found that the
fidelity (probability) is larger than $0.9$ ($0.35$) for
$kx_0\in[0.9,1.03]\pi$ and $\sigma/k$ up to $\simeq 5\%$.}

We now address the dependence of our figures of merit on the
dimensionality of the impurities' spin. While the optimum ratio
$J/v$ depends slightly on $s$, the efficiency of singlet extraction
persists, as shown in Fig.~\ref{Fig2}(c) for
$\rho_{12}\!=\!\ket{s,-s}\bra{s,-s}$ with $s\!=\!1/2,1,3/2$. {Evidently},
$\varrho^{(n)}_{12}$ rapidly converges to the singlet state
regardless of $s$ (for instance, $F^{(n>5)}>0.95$ for $s\!=\!1$) while
$P_{\s\s}^{(n)}$ approaches a \emph{finite} value according to
$P_{\s\s}^{(n\gg{1})}(\rho_{12}){\rightarrow}\left|\bra{\Psi_{s}^{-}}s,-s\rangle\right|^2\!=\!(2s+1)^{-1}$,
exponentially in $n$. Our scheme thus \emph{asymptotically performs
an effective projective measurement onto the spin-$s$ singlet
state}.
As the singlet state has the maximum number of ebits
allowed by the dimension of the Hilbert space of each impurity, the
scheme provides a way to extract more than one ebit by considering
sufficiently high-dimensional impurities' spins. {Moreover, an
entanglement accumulation mechanism is achieved~\cite{mauro-prl}.
For instance, for $s\!=\!2$ and $J/v\!=\!1$ the impurities' entanglement (measured by the
logarithmic negativity, which is upper-bounded by $\log_{2}(d)$ for
a {$d^2$-dimensional} Hilbert space) after $n\!=\!2,4$ and $5$ is
respectively, $1.2,1.8$ and $2$. These are larger than the bound
given by $\log_{2}(2s\!+\!1)$ for $s\!=\!1/2,1$ and $3/2$, making our system
an iteratively exploitable quantum channel: The impurities'
entanglement can be extracted to many pairs of
qubits}~\cite{mauro-prl}. Similar results hold for {any initial
eigenstate of $\hat{S}_{12z}=\hat{S}_{1z}+\hat{S}_{2z}$ with null eigenvalue.

{We now show how the efficiency of singlet-state extraction relies
on resonance-induced selection rules. Let $\ket{s,s,s_{12},m_{12}}$
be the coupled basis of common eigenstates of
$\hat{\mathbf{S}}_{1}^2$, $\hat{\mathbf{S}}_{2}^2$,
$\hat{\mathbf{S}}_{12}^2$ and $\hat{S}_{12z}$ {(the singlet state
thus reads $\ket{\Psi_s^-}\!=\!\ket{s,s,s_{12}\!=\!0,m_{12}\!=\!0}$)}. Let
$\mathcal{E}_{\s}(\rho_{12})$ be the {\it unconditioned} map in
Eq.~(\ref{map_up}) for $\rho_e=\ket{\s}_e\!\bra{\s}$. Clearly, with
the additional {output}-filtering of $\ket{\s}_{e}$,
$\mathcal{E}_{\s}(\rho_{12})$ becomes
$\mathcal{E}_{\s\s}(\rho_{12})$. Notice that in general the product
state $\ket{s,s}_{12}$ is the only fixed point of
$\mathcal{E}_{\s}(\rho_{12})$. However, at resonance ($kx_0\!=\!n\pi$),
$\hat{\mathbf{S}}_{12}^{2}$ is conserved due to the equal
probabilities {of the electron} to be found at each of
$x_i$'s~\cite{ciccarello}. Thus, repeated applications of the
unconditioned map cannot drive the system out of the eigenspace
associated with a set value of $s_{12}$. This and the conservation
of $\hat{S}_{z}$ imply that the singlet state $\ket{\Psi^{-}_{s}}$
becomes an additional fixed point of $\mathcal{E}_{\s}$. Let
$p_{s_{12}}$ be the probability for an injected electron prepared in
$\ket{\s}_e$ to be flipped down when the impurities are prepared in
$\ket{s,s,s_{12},0}$. The selection rules at resonance yield the
evolved impurities'state
$p_{s_{12}}\ket{s,s,s_{12},1}\bra{s,s,s_{12},1}$
$\!+\!(1\!-\!p_{s_{12}})\ket{s,s,s_{12},0}\bra{s,s,s_{12},0}$.} If we
post-select $\ket{\s}_{e}$ at the output ports, each state
$\ket{s,s,s_{12},0}$ with $s_{12}\!\neq\!0$ is left unchanged with
probability $1-p_{s_{12}}$. Under application of
$\mathcal{E}_{\s\s}^{n\gg{1}}$, it thus vanishes as
$(1\!-\!p_{s_{12}})^{n\gg{1}}\!\simeq\! 0$, which clarifies the exponential
convergence exhibited by $F^{(n)}$ and $P_{\s\s}^n$ (cf. Fig.~2).
Differently, $\ket{s,s,s_{12}=0,0}\!=\!\ket{\Psi_{s}^-}$ survives to the
application of $\mathcal{E}_{\s\s}^{n\gg{1}}$ since the selection
rules ensure that $p_{s_{12}=0}=0$~\cite{ciccarello}. If we consider
an element of the uncoupled basis $\ket{\xi}$ such that
$\hat{S}_{12z}\ket{\xi}_{12}\!=\!0$ and expand it over
$\ket{s,s,s_{12},0}$'s, we find that, under application of
$\mathcal{E}_{\s\s}^{n\gg{1}},\ket{\xi}\bra{\xi}\rightarrow\ket{\Psi^{-}_s}$
with a probability $P_{\uparrow\uparrow}^{(n\gg{1})}$ that asymptotically
becomes $|\langle{\Psi}_s^-\vert\xi\rangle|^2$.
When $\ket{\xi}\!=\!\ket{s,-s}_{12}$, as in Fig.~2, the asymptotic
probability is $(2s\!+\!1)^{-1}$.} Our clear interpretation of
the physics behind our protocol is an important feature for the
development of novel schemes.

Unlike previous
proposals~\cite{yang-yasser-giorgi,ciccarello,nakazato}, a
remarkable advantage of our protocol is that it can be applied to
magnetic impurities of spin higher than 1/2. {For instance, we could
use a 1D semiconducting wire with embedded} Mn impurities having
$s\!=\!5/2$. Although impressive progresses have been made, a major
obstacle in spintronics implementations is the current lack of
high-efficiency electron-spin filters~\cite{spintronics}. As a way
to overcome such difficulties, we discuss an alternative system [see
Fig.~\ref{Fig1}(b)] able to act as an accurate simulator of
$\hat{H}$ and {holding} the promises for not far-fetched
experimental implementation. The basic idea is to replace the
electron with a single photon propagating in a 1D photonic waveguide
sustaining two frequency-degenerate orthogonally polarized modes.
{For consistency of notation, we denote circular polarizations by
$\s$ and $\g$}. Each impurity is now embodied by a multilevel atom
[see Fig.~\ref{Fig1}(c)] having a $(2s+1)$-fold degenerate ground
level spanned by $\{\ket{g_{-s}},..,\ket{g_{s}}\}$ and a $2s$-fold
degenerate excited level spanned by
$\{\ket{e_{-s}},..,\ket{e_{s-1}}\}$. The standard three-level
$\Lambda$ and five-level M configurations are recovered, for
instance, by taking $s\!=\!1/2$ and $s\!=\!1$, respectively. {Such a
configuration may be found in the rich hyperfine spectrum of alkali
atoms}. We assume electric-dipole selection rules such that each
$\ket{e_{m}}$ ($m\!=\!-s,..,s-1$) is connected to the pair of
nearest-neighbor ground states $\{\ket{g_m},\ket{g_{m\!+\!1}}\}$ via
coherent scattering of a photon between the two orthogonally
polarized modes. To fix the ideas, we take the transition
$\ket{e_m}\leftrightarrow\ket{g_{m}}$
($\ket{e_m}\leftrightarrow\ket{g_{m\!+\!1}}$) to be driven by the
$\s$-polarized ($\g$-polarized) mode. Each atom can thus undergo a
transition between two adjacent ground states
$\ket{g_m}\leftrightarrow\ket{g_{m\!+\!1}}$ via a two-photon Raman
process with associated coherent scattering of a photon between
states $\ket{\s}$ and $\ket{\g}$. Assuming a linear dispersion law
$E=v_{ph}k$ with $v_{ph}$ the group velocity of the photon and $E$
its energy, the free Hamiltonian of the field in the waveguide
is~\cite{shen_fan}
$\hat{H}_{ph}\!=\!-i\sum_{\beta\!=\!R,L}\sum_{\gamma=\s,\g}\int dx\,
v_{\beta}\,\hat{c}_{\beta,\gamma}^\dag(x){\partial}_x\hat{c}_{\beta,
\gamma}(x)$ with $v_{R}\!=\!-v_{L}=v_{ph}$ and
$\hat{c}_{R,\gamma}^\dag(x)$ [$\hat{c}_{L,\gamma}^\dag(x)$] the
bosonic operator creating a right (left) propagating photon of
polarization $\gamma$ at position $x$. Considering dipole
transitions with Rabi frequencies and natural excited-state
linewidth smaller than the corresponding detuning from the excited
state, each state $\ket{e_{m}}$ is only virtually populated and the
effective atom-photon coupling reads
$\hat{V}\!=\!\sum_{i\!=\!1,2}\!\int{d}x(\hat{c}_{\s}^\dag(x)\hat{c}_{\g}(x)\hat{\mathcal{S}}_{i-}\!+\!\mathrm{h.c.})\,\delta(x\!-\!x_i)$
with $c_{\gamma}^\dag(x)\!=\!\sum_{\beta\!=\!R,L}c_{\beta,\gamma}^\dag(x)$
and
$\hat{\mathcal{S}}_{i+}\!=\!\hat{\mathcal{S}}^\dag_{i-}\!=\!\sum^{s-1}_{m\!=\!-s}J_{s,m}\ket{g_{m\!+\!1}}_i\bra{g_{m}}$.
{Here $J_{s,m}$ is the effective transition rate of the Raman
process leading the $i$-th atom from $\ket{g_m}_i$ to
$\ket{g_{m\!+\!1}}_i$, assuming identical atoms.} We map the photonic
polarization into an effective pseudospin-$s$ as
$\hat{\mbox{\boldmath$\sigma$}}\!=\!\int
dx\,\hat{\mbox{\boldmath$\sigma$}}(x)$ with
$\hat{\sigma}_+(x)=\hat{\sigma}^\dag_-(x)=c_{\s}^\dag(x)c_{\g}(x)$
and
$\hat{\sigma}_z(x)=[\hat{c}_{\s}^\dag(x)\hat{c}_{\s}(x)-\hat{c}_{\g}^\dag(x)\hat{c}_{\g}(x)]/2$.
{Provided that $J_{s,m}=J \,\chi_{s,m}$ with
$\chi_{s,m}=[s(s+1)-m(m+1)]^{1/2}$, each $\hat{\mathcal{S}}_{i\pm}$
becomes the effective pseudospin-$s$ operator
$\hat{\mathcal{S}}_{i\pm}=J\hat{S}_{i\pm}$, where $\hat{S}_{i\pm}$
obeys the standard algebra of angular momentum. Under these
conditions, this model can be regarded as the second quantization
version of $\hat{H}$ with the exchange electron-impurity coupling
replaced by an isotropic XY interaction. It is easily checked that
$[\hat{H}_{ph}+\hat{V},\hat{S}_z]=0$ and, provided $kx_0/\pi\in
\mathbb{Z}$,} $[\hat{H}_{ph}+\hat{V},\hat{\mathbf{S}}_{12}^2]=0$.
Through standard procedures~\cite{shen_fan}, we have derived the
stationary states $\ket{k,m'_{ph},\{m'_i\}}$ for a single photon with
wavevector $k$ ($m'_{ph}$ is the quantum number of $\hat{\sigma}_z$).
The input (output) part of $\ket{k,m'_{ph},\{m'_i\}}$ is
formally analogous to $\ket{k,m'_{e},\{m'_i\}}_{in}$ ($\ket{k,m'_{e},\{m'_i\}}_{out}$). Here, $\mathcal{E}_{\s\s}(\rho_{12})$ is
obtained analogously to what is done for the previous model with
photonic polarization detection used for the post-selection. Plots
analogous to {those in} Figs.~\ref{Fig2} are reproduced with only
negligible quantitative differences. Practically,
$\mathcal{E}_{\s}(\rho_{12})$ is obtained using Geiger-like
photodetectors at the input/output ports of the waveguide combined
with polarizing beam-splitters to realize
$\mathcal{E}_{\s\s}(\rho_{12})$. {Each $J_{s,m}$ depends on the
product of the Clebsch-Gordan coefficients associated with the
far-detuned (one-photon) transitions involved in the process
$\ket{g_m}\leftrightarrow\ket{g_{m+1}}$. The condition $J_{s,m}=J
\,\chi_{s,m}$ is clearly fulfilled for $s\!=\!1/2$, involving only
$\chi_{1/2,-1/2}\!=\!1$. For $s\ge1$ the pattern of $J_{s,m}$'s might in
general deviate from the ideal one dictated by the $\chi_{s,m}$'s.
However, we have assessed $F^{(n)}$ and $P^{n}_{\s\s}$ finding that our scheme is strikingly
robust against such deviations~\cite{inpreparation}. {For instance,
for $s\!=\!3/2$, the ideal pattern yields $J_{3/2,1/2}/J_{3/2,-3/2}\!=\!1$
and $J_{3/2,-1/2}/J_{3/2,-3/2}\!=\!2/\sqrt{3}$. By taking
$J_{3/2,-3/2}/\,v_{ph}\!=\!J_{3/2,1/2}/\,v_{ph}=\sqrt{3}$ and
$J_{3/2,-1/2}/\,v_{ph}\!=\!4\sqrt{3}$, which are far from ideal, we
obtain $F^{(n>6)}\!=\!0.97$, and $P^{(n>6)}\!=\!0.26$. These values are
basically identical to the values obtained with the ideal ratios.
This alternative model turns out to be also robust against
deviations of $k$ from the ideal resonance
conditions~\cite{inpreparation}. Our protocol is thus resilient and
flexible to the actual working conditions.}

For a realization of the scheme {in the case $s\!=\!1/2$},
the impurities can be embodied by $\Lambda$ configurations
encompassed in the (single-electron charged) trionic picture of
semiconducting quantum dots {(QDs)}, which have been the center of
extensive studies~\cite{imamoglu}. Positioning QDs within a
waveguide or a cavity is now achievable with high accuracy
($\sim\!{30}$nm). It can be easily shown that for
a photonic wavelength of $780$nm in a GaAs structure ($400$nm in a
GaN nanowire), $x_0\sim{0.1}\mu\!$m (${1}\mu\!$m) is required for
the resonance condition, which is achievable. Strong coupling
between a single QD and a cavity field has been
demonstrated~\cite{imamoglu} and current experimental efforts
make the achievement of $J/v\sim{1}$ realistic in large
refractive-index structures, without the need of waveguide's bandgap. {We consider~GaInN (InAs) QDs in GaN (GaAs)
nanowires as potential candidates for our scheme.} Their typical
quality factor is~$\simeq\!{10}^3$, implying single-photon lifetime
$\tau_p\sim{1}$ps at ${4}00$nm wavelength. The
refractive index of GaN is $\sim2$, so that a photon travels
$x_0\!=\!1\mu$m in ${\tau_p}/100$.
Ongoing experimental progresses make the controlled growth and positioning of two
QDs in $\mu$m-long waveguides, quite
realistic.

We have proposed a scheme for the conditional extraction of singlet
states of two remote {spin$\!-\! s$'s} based on projective measurements
{over interaction mediators.} The protocol does not require the
demanding recycling of the same mediator.
It {achieves $s\!+\!1/2$ ebits} with finite probability, a small number
of steps, weak requirements on the parameters entering the dynamics
and no interaction-time tuning. We have proposed a realistic setup
where the mediators are embodied by photons and the spins to be
entangled by artificial atoms.

We thank M. Weber, G. Fishman, F. Julien, J.-M. Lourtioz, Y. Omar,
R. Passante, L. Rizzuto and M. Tchernycheva. We acknowledge support from PRIN 2006 ``Quantum noise
in mesoscopic systems'', The Leverhulme Trust, EPSRC, QIPIRC and the
British Council/MIUR British-Italian Partnership Programme
2007-2008.


\begin {thebibliography}{99}
\bibitem{varie1} D.N. Matsukevich, {\it et al.}, \prl {\bf 96}, 030405
(2006); B. Julsgaard {\it et al.}, Nature (London) {\bf 432}, 482
(2004); J. Volz {\it et al.}, \prl {\bf 96}, 0304004 (2006).
\bibitem{varie2} M. Paternostro, W. Son, and M.S. Kim, \prl {\bf 92},
197901 (2004).
\bibitem{yang-yasser-giorgi} D. Yang, S.-J. Gu, and H. Li,
quant-ph/0503131; A.T. Costa, Jr., S. Bose, and Y. Omar, Phys. Rev.
Lett. \textbf{96}, 230501 (2006); G.L. Giorgi and F. De Pasquale,
Phys. Rev. B \textbf{74}, 153308 (2006).
\bibitem{ciccarello}  F. Ciccarello \emph{et al.}, New J. Phys. {\bf 8},
214 (2006); J. Phys. A: Math. Theor. \textbf{40}, 7993 (2007); Las.
Phys. \textbf{17}, 889 (2007); F. Ciccarello, G.M. Palma, and M.
Zarcone, Phys. Rev. B \textbf{75}, 205415 (2007)
\bibitem{nakazato}  K. Yuasa and H. Nakazato, J. Phys. A: Math. Theor.
\textbf{40}, 297 (2007).
\bibitem{smer}H. Nakazato, M. Unoki, and K. Yuasa,~\pra {\bf 70}, 012303
(2004); L.-A. Wu, D.A. Lidar, and S. Schneider, {\it ibid.}~{\bf
70}, 032322 (2004);  G. Compagno \emph{et al.}, {\it
ibid.}~\textbf{70}, 052316 (2004).
\bibitem{mauro-prl} M. Paternostro, M.S. Kim, and G.M. Palma, Phys. Rev.
Lett. \textbf{98}, 140504 (2007).
\bibitem{daniel} For quantum state engineering via iterated quantum
operations, see D. Burgarth and V. Giovannetti,
New J. Phys. 9, 150 (2007).
\bibitem{datta} S.~Datta, Electron~Transport~in~Mesoscopic~Systems
(Cambridge University Press, Cambridge, 1997).
\bibitem{nanotube} S. J. Tans \emph{et al.}, Nature (London) \textbf{386},
474 (1997).
\bibitem{marsiglio} For the single-impurity case {without spin-flip} see
W. Kim, R.K. Teshima and F. Marsiglio, Europhys. Lett. \textbf{69},
595 (2005).
\bibitem{spintronics} D.D. Awschalom, D. Loss, and N. Samarth,
\textit{Semiconductor~Spintronics~and~Quantum Computation}
(Springer, Berlin, 2002).
\bibitem{shen_fan} J.-T. Shen and S. Fan, Phys. Rev. Lett. \textbf{95},
213001 (2005); {\it ibid.} \textbf{98}, 153003 (2007).
\bibitem{inpreparation} F. Ciccarello \emph{et al.}, in preparation.
\bibitem{imamoglu} M. Atat\"ure {\it et al.}, Science {\bf 312}, 551
(2006); K. Hennessy, {\it et al.}, Nature (London) {\bf 445}, 896
(2007).
\end {thebibliography}
\end{document}